%% file: majorana_prox.tex
\documentclass[prl,twocolumn,floatfix]{revtex4-1}

\usepackage{amsmath,amssymb,wscmath}
\usepackage{color,framed}
\usepackage[pdftex]{hyperref,graphicx}
\hypersetup{colorlinks = true, urlcolor = blue, linkcolor = blue, citecolor = blue}

\begin{document}
\title{Proximity effect and Majorana bound states in clean semiconductor nanowires coupled to disordered superconductors}
\author{William S. Cole, Jay D. Sau, S. Das Sarma}
\affiliation{Condensed Matter Theory Center and Joint Quantum Institute,
Department of Physics, University of Maryland, College Park, MD 20742, USA}
\begin{abstract}
We study a semiconductor wire with strong spin-orbit coupling, which is proximity-coupled to a superconductor with chemical potential disorder. When tunneling at the semiconductor-superconductor interface is very weak, it is known that disorder in the superconductor does not affect the induced superconductivity nor, therefore, the effective topological superconductivity that emerges above a critical magnetic field. We demonstrate nonperturbatively how this result breaks down with stronger proximity coupling by obtaining the low-energy (i.e., subgap) excitation spectrum through direct numerical diagonalization of an appropriate BdG hamiltonian.
With strong proximity coupling, we find that disorder in the parent superconductor suppresses the (non-topological) induced gap at zero magnetic field by disordering the induced pair potential. In the topological superconducting phase at large magnetic field, strong proximity coupling reduces the localization length of Majorana bound states, such that the induced disorder eliminates the topological gap while \emph{bulk} Majorana zero modes emerge, even for short wires.
\end{abstract}
\date{\today}
\maketitle


Topological superconductors (TS), though exceedingly rare among naturally occuring superconducting materials, can be synthesized through the proximity effect between a conventional superconductor (SC) and either the surface of a topological insulator (TI) \cite{fu_kane_tsc} or a semiconductor (SM) with strong spin-orbit coupling and spin splitting \cite{sau_rashba_2d_prl}. Presently, most solid state experiments follow this ``recipe" (similar ideas for producing topological superfluids have been proposed for ultracold atomic systems \cite{zhang_prl2008, sato_prl2009}, though we restrict our attention here to solid state systems only). After several experimental groups reported promising measurements consistent with TS \cite{mourik-expt, das-expt, deng-expt, finck-expt, churchill-expt, rokhinson_frac_ac}, lingering disagreements between theory and experiment motivated the development of an increasingly sophisticated theoretical picture. This has allowed some controversy to persist as to whether or not TS has truly been observed.

The earliest proposals for this kind of synthetic TS, following \cite{fu_kane_tsc}, modeled induced superconductivity under the assumption that a uniform $s$-wave pair potential may be added to a non-superconducting hamiltonian (e.g., that of a TI surface) and that any further effects of coupling to the SC can be ignored. This approximation turns out to be qualitatively, and even quantitatively, accurate for ``weak" proximity coupling \cite{sau_robust, tudor_prox}. Explicitly, if the proximity coupling is characterized by an energy scale $\gamma$, while the parent superconductor gap is $\Delta_{sc}$, then this model is viable when $\gamma/\Delta_{sc} \ll 1$. Unfortunately for applications, the induced spectral gap in this regime is approximately $\gamma$, while a large induced gap (relative to the energy resolution, temperature, etc., of the specific experiment) is a necessary requirement for experimentally robust signatures of in-gap modes. An additional requirement is that the proximity induced gap be \emph{hard}, that is, free of sub-induced-gap states that contribute to low-bias conductance. This in turn demands a uniform SC-nanowire coupling along the length of the wire \cite{takei_prl}.

In an effort to fabricate TS devices with large and hard induced gaps, epitaxial superconductor/semiconductor hybrids have recently been produced and characterized in this strong proximity coupling regime \cite{cm_natnano}
(a similar platform relies on metal-on-superconductor hybrids which are \emph{intrinsically} in this regime \cite{brydon_prb, yazdani_science}.) It is, therefore, worth understanding the signatures of deviations from this simplified model in the context of the intense current experimental activity \cite{cm_nature_exppro} for the laboratory realization of synthetic TS based on proximity-induced superconductivity. There is a growing consensus that the consequences of strong proximity coupling are crucial for accurate theoretical interpretation of experiments. 

One such implication, however, has received little attention. To lowest order in $\gamma/\Delta_{sc}$, the induced superconductivity is known to be ``protected" against disorder in the \emph{parent} superconductor \cite{lutchyn_disorder, potter_lee_erratum}. This is a reassuring result - even a very dirty parent superconductor will not introduce disorder into the effective low-energy TS description of the wire.  At intermediate or strong proximity coupling, though, this perturbative result must break down. Along with any superconducting correlations, strong coupling to a disordered superconductor must also give rise to effective disorder in an otherwise perfectly clean system. This work seeks to characterize the combined nonperturbative effects of strong proximity coupling and parent superconductor disorder.

We begin our study with a relatively minimal model of a single-channel, clean, ballistic nanowire with spin-orbit coupling and Zeeman splitting with a large $g$ factor \cite{lutchyn_wire, oreg_wire}. This wire is coupled through a uniform interface hopping to a conventional $s$-wave superconducting bulk (represented by a ribbon, with width much greater than the SC coherence length). Our goal is to treat the composite wire and SC system in a nonperturbative, approximation-free manner to the extent that the wire is noninteracting and the parent superconductor can be treated within the BdG mean-field approximation. Because the model is fundamentally quadratic, it seems that it should be amenable to an exact numerical solution. However, there are length scale and energy scale considerations that make a full diagonalization unreasonable \emph{in practice}, which has generally motivated the use of simpler effective models. For semiconductor/superconductor hybrids (which are the most experimentally relevant systems), the Fermi wavelength of electrons in the semiconductor wire is typically of the order of tens of nm, while the Fermi wavelength of electrons in the superconductor is a fraction of a nm. A typical superconducting coherence length could be $\sim \mu$m for Aluminum or tens of nm for Niobium. Additionally, the wire is expected to have a long mean-free path, while the superconductor mean-free path is typically significantly shorter than the coherence length.

In the following, we will \emph{not} pursue a detailed quantitative simulation of any existing experiments, rather a qualitative approach in which this hierarchy of length scales can be captured without making any additional approximations that may obscure the effects of induced disorder. We have overcome the practical limitation of diagonalizing a sufficiently large hamiltonian to capture these widely varying scales by abandoning the goal of such a full diagonalization outright. We are only interested in a relatively small subset of states with energy eigenvalues below (or on the order of) the parent SC gap, which is itself a much smaller energy scale than the band structure scales (i.e., chemical potential in the SC and hopping matrix elements). With this understanding, we can leverage the sparse matrix structure of the problem and use the Lanczos method, with the shift-and-invert scheme appropriate to interior eigenvalue problems, to resolve only the part of the spectrum corresponding to the induced (topological) superconductivity, while using very large lattices. More details on the model and method of solution are relegated to the supplementary material.

Our main findings are the following:
(1) We introduce nonmagnetic (specifically, chemical potential) disorder in the SC, first without any external magnetic field. There is no intrinsic disorder in the wire. For sufficiently strong interface coupling we observe ``pair-breaking" behavior in the \emph{induced} superconductivity, despite the explicit presence of time reversal symmetry, in apparent violation of Anderson's theorem \cite{anderson_thm}. We attribute this to a disordered SC-induced pair potential in the wire.
(2) As even moderate chemical potential disorder is expected to be catastrophic for the realization of \emph{isolated} Majorana zero modes (MZMs) in the TS phase of the wire \cite{motrunich_pwave}, we also consider the combined effect of SC disorder and an external magnetic field. The topological gap is \emph{extremely} fragile against disorder at strong proximity coupling, corresponding to a proliferation of near-zero energy localized states, which produce a Griffiths-like singularity \cite{motrunich_pwave} in the density of states. Observation of non-Abelian statistics would be essentially impossible in this Griffiths phase, although some robust signatures of MZMs (e.g. zero bias conductance peaks) could still be present.


\begin{figure}[t]
\centering
\includegraphics[width=0.48\textwidth]{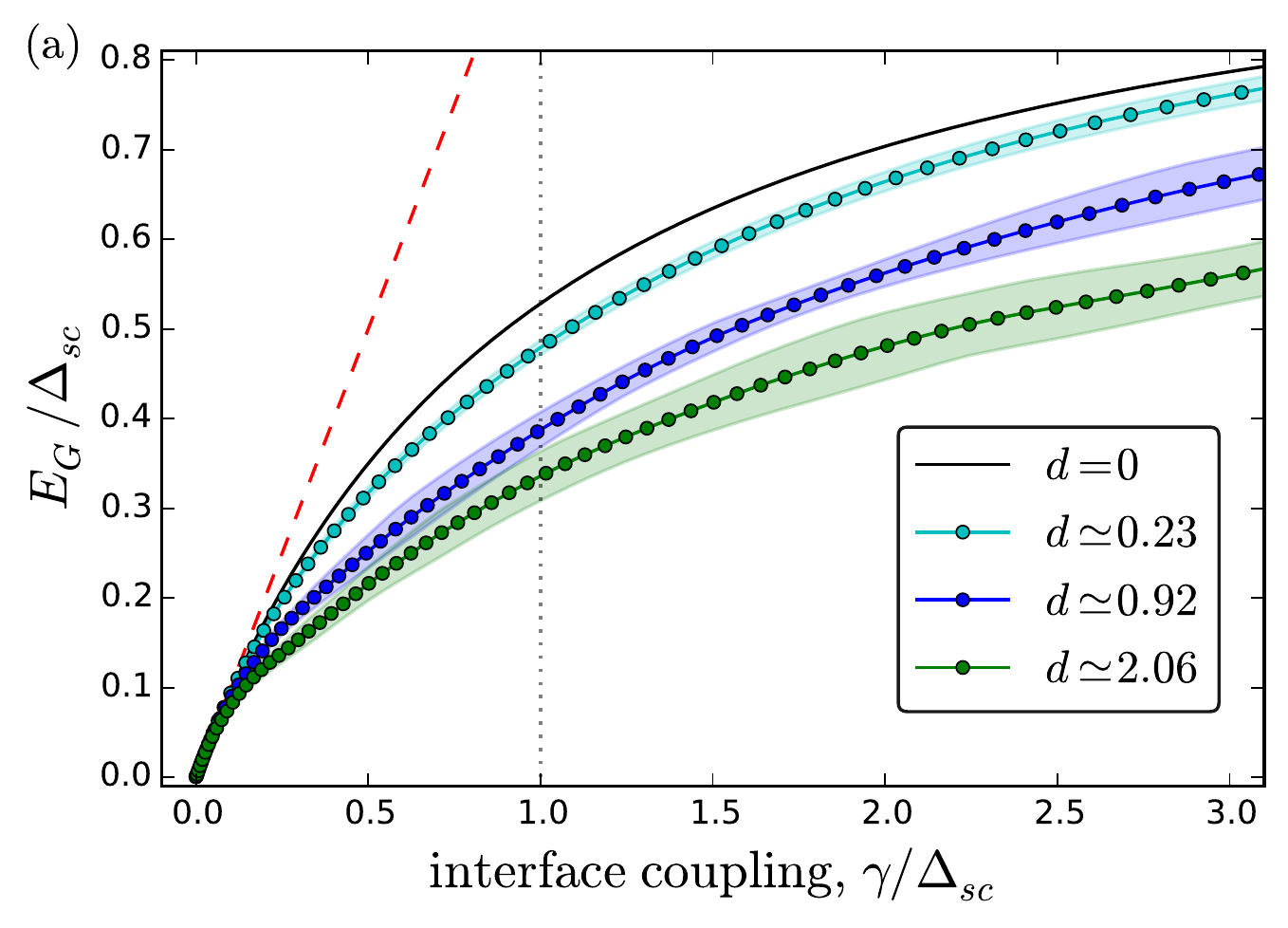}
\includegraphics[width=0.48\textwidth]{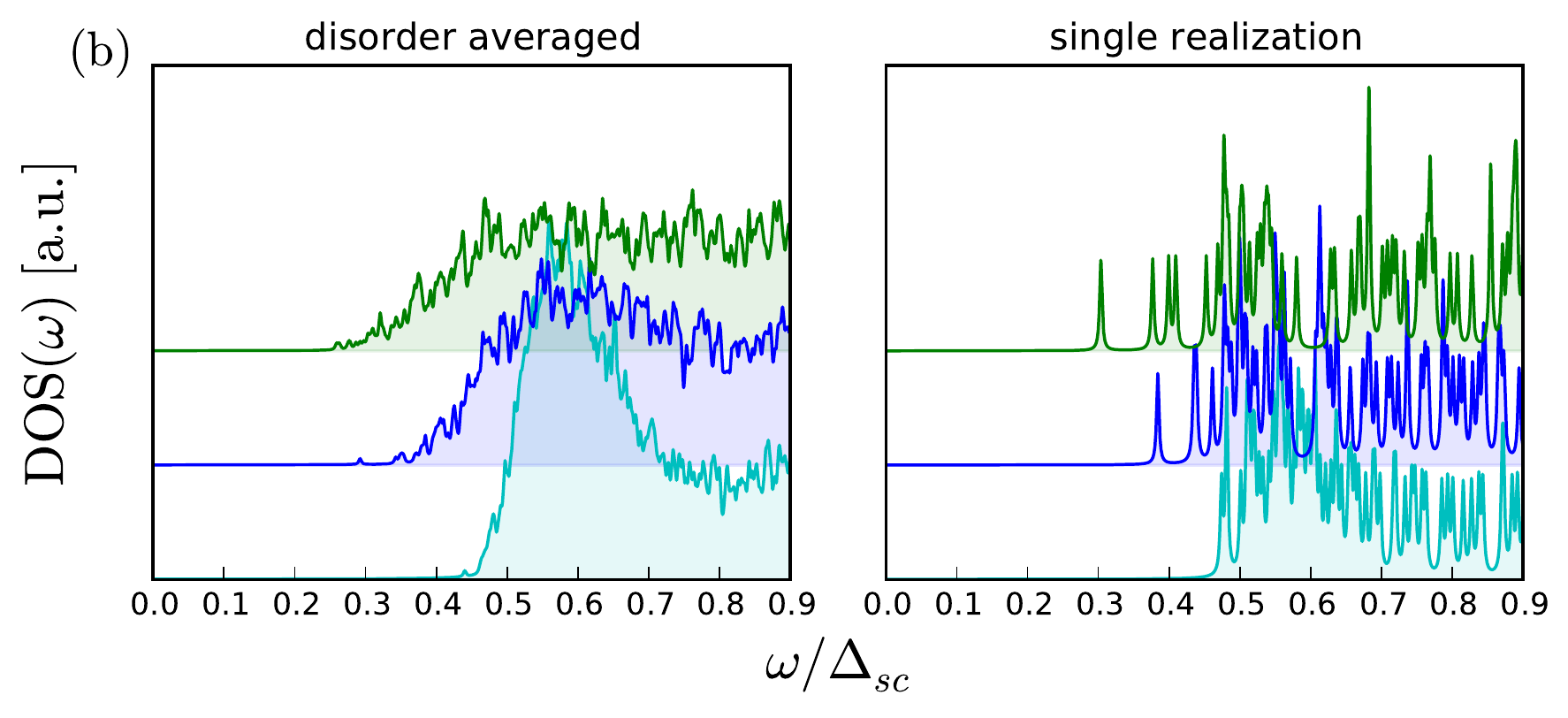}
\caption{
(Color online)
(a) Induced gap (in units of parent superconductor gap) versus interface coupling for several values of disorder. We plot this over a moderate range of coupling, showing that (i) disorder suppresses the induced gap substantially at intermediate coupling, $\gamma/\Delta_{sc} \simeq 1$, while (ii) for sufficiently small $\gamma$, all of the curves recover the expected $E_G = \gamma$ limit (red dashed line) as $\gamma \rightarrow 0$. In this limit the wire is ``protected" from bulk SC disorder.
(b) Disorder averaged and single realization subgap densities of states at intermediate coupling ($\gamma/\Delta_{sc} \simeq 1$). Each disorder averaged curve is obtained from the same set of 30 independent samples. Disorder in the parent SC broadens the subgap ``coherence peak" and produces bound states below the clean-limit induced gap (thus suppressing $E_G$).
}
\flabel{1}
\end{figure}

{\bf Spectral gap at zero field:}
First, we calculate the excitation gap $E_G$, which is the smallest eigenvalue of the full (wire + SC bulk) hamiltonian, as a function of the interface coupling $\gamma$ and a dimensionless measure of the disorder strength $d$ in the superconductor (see supplement). In the limit of zero disorder and zero magnetic field, this also coincides with the effective induced pair potential, although the two scales are otherwise generally unrelated. Our exact numerical results for $E_G$, shown in \fref{1}(a) in the absence of spin splitting, agree qualitatively with a recent self-consistent Born approximation (SCBA) calculation \cite{hoi_SCBA}. We recover the perturbative result \cite{lutchyn_disorder} that SC disorder does not affect $E_G$ for $\gamma/\Delta_{sc} \ll 1$, however, at stronger coupling disorder does substantially suppress the induced gap (although it has no effect on the gap of the \emph{parent} SC, which is $\Delta_{sc}$ regardless of the disorder by virtue of Anderson's theorem). Fluctuations around an averaged $E_G$ arise from sampling several independent disorder realizations. Vanishing sample-to-sample fluctuations for $\gamma/\Delta_{sc} \ll 1$ further confirm the disorder independence of $E_G$ in that weak proximity coupling regime, however, these fluctuations do become increasingly dramatic with disorder at moderate interface coupling, $\gamma \simeq \Delta_{sc}$.

In \fref{1}(b), we consider a cut at fixed $\gamma = \Delta_{sc}$ and show characteristic densities of states (averaged over 30 disorder realizations) for several values of $d$. In the disorder-free case, the subgap spectrum exhibits a ``coherence peak" of states accumulating at $E_G$. However, even for weak disorder, this peak is broadened with the lowest-lying states shifting to energies well below the clean $E_G$. These are phenomenological signatures of pair breaking. As an alternative perspective, we also show the DOS evaluated for a \emph{single} disorder realization. Here, we can see that the high-energy states (that is, near the parent gap) form a broad continuum, while the lowest energy states (which determine $E_G$) are energetically isolated and correspondingly spatially localized. Because of this localization, these states will not contribute to the wire edge LDOS, probed via the conductance of a normal metal-superconductor junction, unless by accident they are within a localization length of the wire edge.

For zero field, this suppression of the induced gap and pair-breaking behavior in the DOS is an initially surprising result in light of Anderson's theorem, which suggests that conventional $s$-wave superconductors are immune to disorder that preserves time-reversal symmetry (TRS) \cite{anderson_thm}. We indeed find that the \emph{parent} SC gap remains constant despite the presence of disorder. This apparent pair breaking effect is limited to the induced superconductivity in the wire.

It is not universally realized that Anderson's theorem actually relies both on TRS \emph{and} an approximately spatially uniform pair potential $\Delta(\vec{r}) \simeq \Delta$. An inhomogeneous pair potential also can produce bound states and give rise to pair breaking, even in the presence of TRS. In conventional superconductors with a small gap and a correspondingly large correlation length, a self-consistent pair potential disorder is energetically costly, so this physics is not relevant even for very dirty samples of, say, Aluminum or Niobium on their own. There is no such penalty for a strongly inhomogeneous \emph{induced} pair potential in a hybrid device with no intrinsic attractive interaction in the semiconductor, as is clear from our numerical calculation despite a spatially uniform wire-SC coupling and uniform pair potential in the SC. This inhomogeneous proximity-induced pair potential leads to a ``violation" of Anderson's theorem in the clean wire, even without any spin-orbit coupling or spin splitting (i.e. even in the non-topological $s$-wave SC regime). This is qualitatively similar to an effective model for pair-potential disorder studied previously in this context \cite{takei_prl}, although the mechanism is quite different. Here, the pairing inhomogeneity appears even though the SC-wire interface is perfectly smooth.


\begin{figure}[t]
\centering
\includegraphics[width=0.48\textwidth]{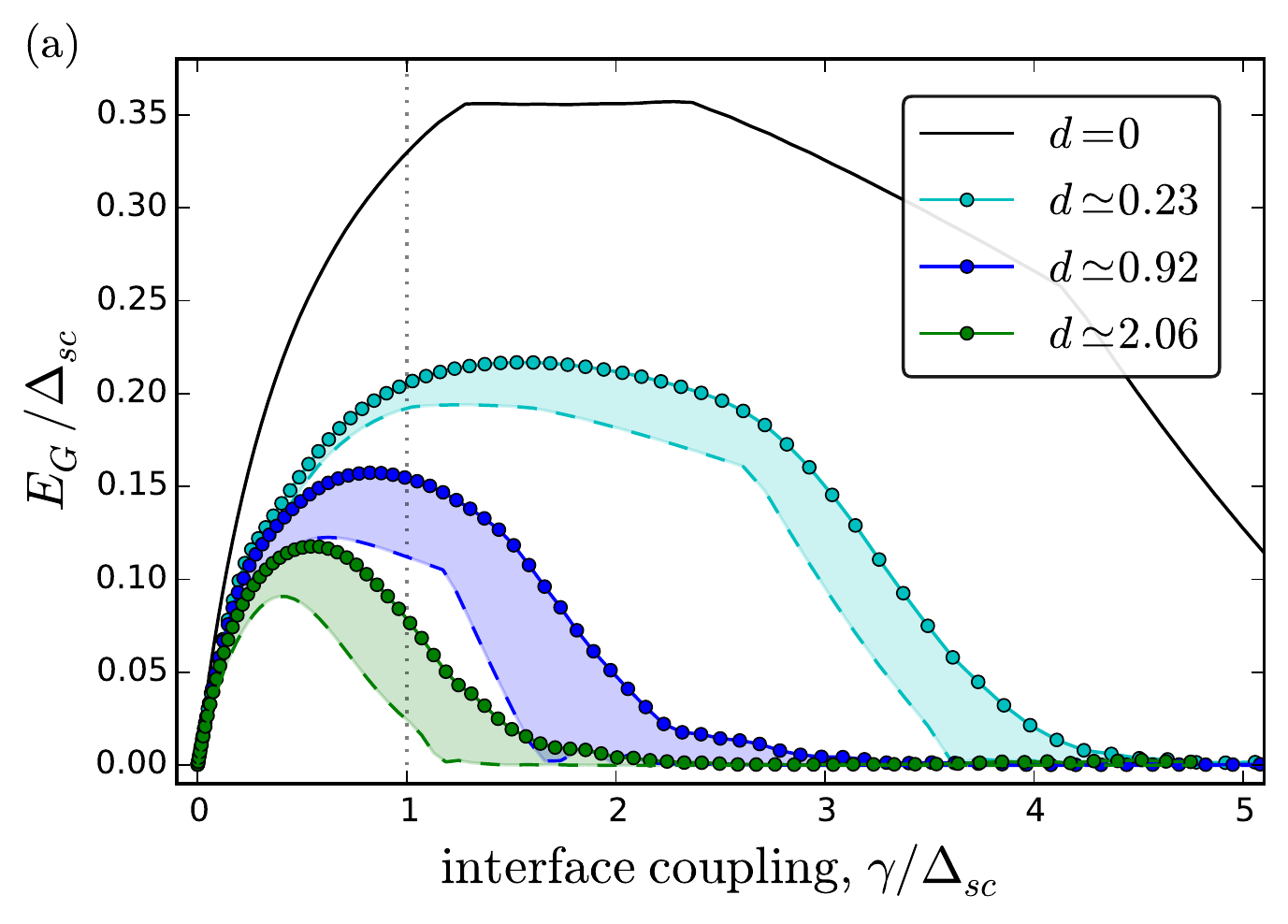}
\includegraphics[width=0.48\textwidth]{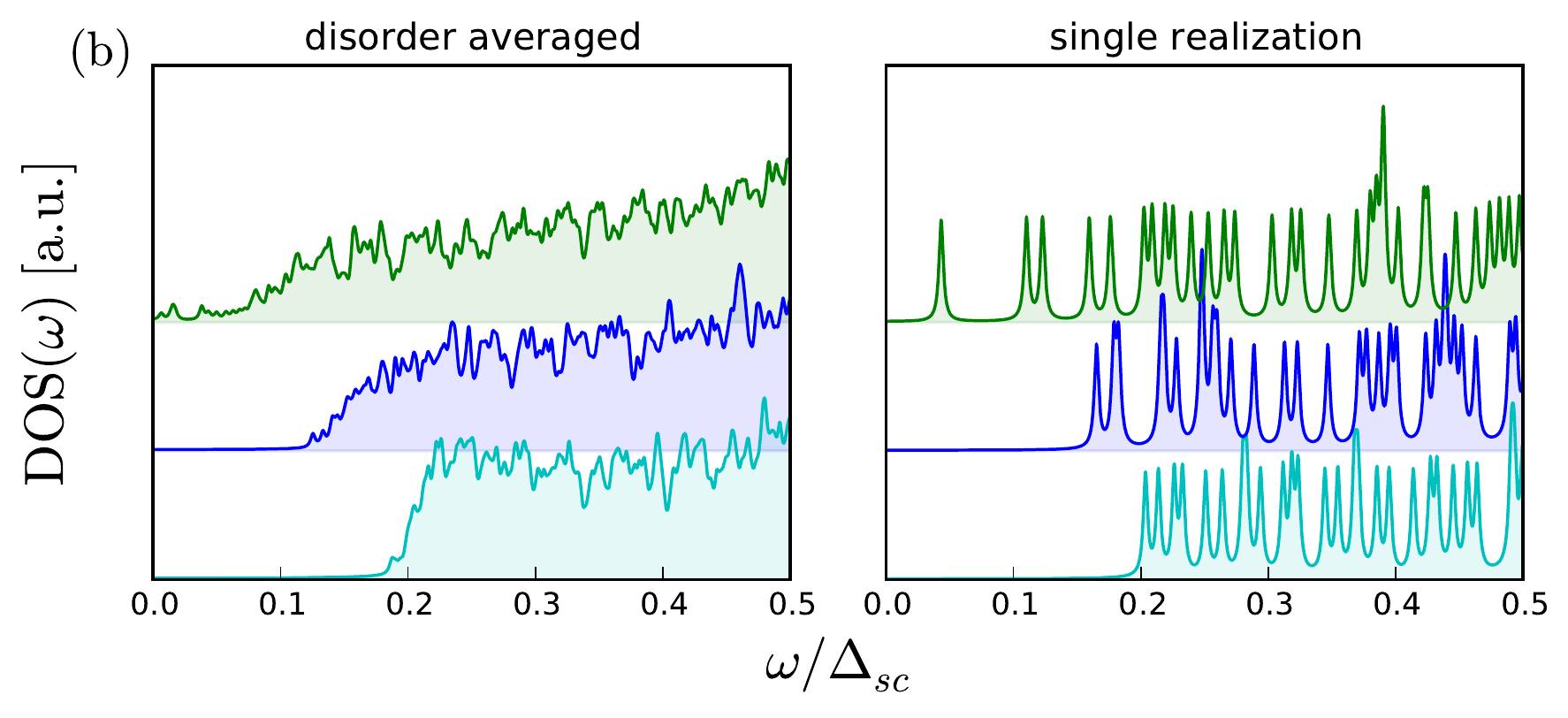}
\caption{
(Color online)
(a) Induced \emph{topological} gap versus interface coupling energy for several values of disorder, with a bare wire Zeeman energy $V^Z_{nw} = 4 \Delta_{sc}$, such that the wire is topological over the entire coupling window in the clean limit. Compared to \fref{1}, the gap suppression is much more substantial. The topological gap however also exhibits ``protection" from bulk SC disorder at small $\gamma$. The data points correspond to disorder averaging, while the dashed lines designate the \emph{minimum} $E_G$ over all sampled disorder realizations.
(b) Disorder averaged subgap densities of states for several representative values of disorder at intermediate coupling ($\gamma/\Delta_{sc} \simeq 1$), as in \fref{1}. For sufficiently strong disorder, the DOS acquires a long tail such that the disorder averaged DOS closes, even though most individual realizations have a noticeable gap.
}
\flabel{2}
\end{figure}

\begin{figure}[t]
\centering
\includegraphics[width=0.48\textwidth]{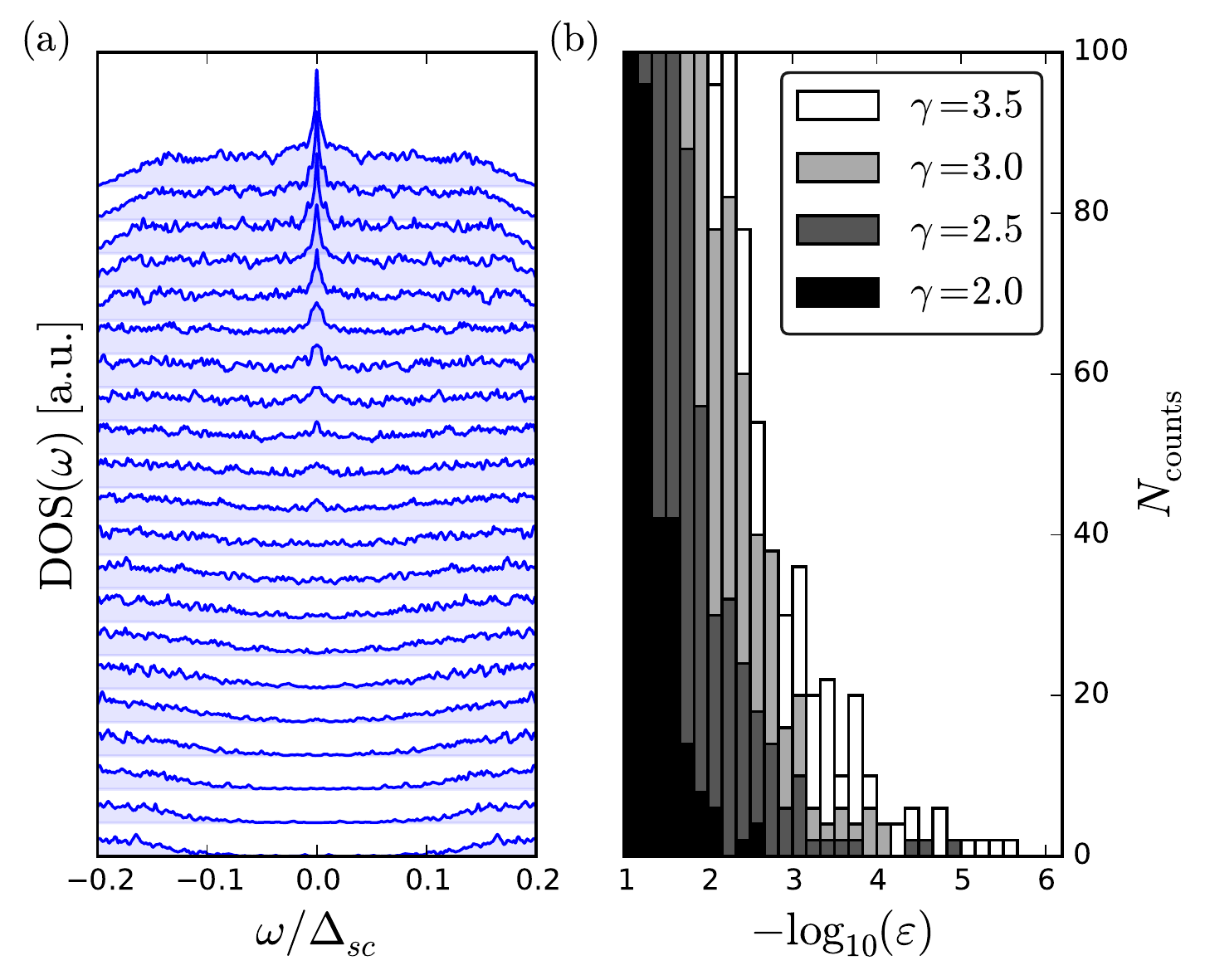}
\caption{
(Color online)
(a) Disorder averaged low-energy DOS for $d\simeq0.92$ with increasing $\gamma=1.5-3.5 \Delta_{sc}$ from bottom to top, demonstrating the closure of the gap and eventual emergence of a singularity at zero-energy.
(b) Distribution of the negative logarithm of low-energy eigenvalues for several values of coupling. Exponentially small eigenvalues are exponentially rare, however the \emph{relative} probability of obtaining an exponentially small eigenvalue increases substantially with $\gamma$.
}
\flabel{3}
\end{figure}

{\bf Spectral gap in the topological regime:}
Next, we can proceed with a similar analysis for the effect of SC disorder on the subgap spectrum with a very large spin splitting in the wire, $V^Z_{nw} = 4 \Delta_{sc}$. In a naive approximation, topological superconductivity is achieved (in the clean limit) when $V^Z_{nw} > \gamma$ \cite{chevallier_disorder_coupling}. The ``topological gap" cannot be a monotonically increasing function of $\gamma$, since increasing $\gamma$ in turn reduces the effective spin splitting, so crossing $\gamma > V^Z_{nw}$ drives a gap-closing transition back to the trivial state. In our numerical results, this actually occurs at a larger value of $\gamma$, attributable to including the small, typically neglected, Zeeman energy in the SC.

In \fref{2}(a), we show the disorder-averaged excitation gap $E_G$, with the dashed lines here highlighting the \emph{minimum} value of $E_G$ over several disorder realizations. (We have used periodic boundary conditions to avoid the contribution from edge MZMs.) We emphasize that these are, of course, both length-dependent quantities. In the limit of an infinite length of wire both of these measures must coincide and tend to zero. A finite wire length and fixed number of disorder samples is, however, reflective of the distribution of gap measurements from a finite set of devices in experiment. That the average $E_G$ is substantially larger than the minimum $E_G$ suggests that the finite system disorder-averaged DOS is characterized by a long tail of low-energy but unlikely states, as verified in \fref{2}(b).

The general fragility of the underlying 1D effective (topological) $p$-wave superconductivity against disorder is well understood \cite{motrunich_pwave}, however, the specific dependence on the proximity coupling is interesting and has not been studied before. For any value of disorder, at sufficiently strong coupling the average $E_G$ in \fref{2}(a) vanishes. That is, $E_G \sim 0$ regardless of the particular disorder configuration. However, this closing is not sharp. For each $d$, the average $E_G$ shows a ``foot" well after the minimum $E_G$ has gone essentially to zero. This is the Griffiths effect - in this crossover window, any \emph{individual} disorder realization will most likely have a comparably large $E_G$, but in some ``rare" disorder configurations (or, equivalently, for a sufficiently long wire) the smallest $E_G$ is exponentially small.

To explore this further, in \fref{3}(a), we track the disorder averaged density of states for $d \simeq 0.92$ along this foot, from $\gamma/\Delta_{sc} = 1.5-3.5$. We see that the gap in the DOS closes and is replaced by a Griffiths singularity at zero energy, as every disorder realization provides near-zero energy states.

This Griffiths effect provides a useful alternative explanation of \fref{2}(a). The probability of the disorder effectively producing a segment of wire (topological embedded in non-topological or vice versa) with two MZMs separated by a length $L$ is $P(L) \propto e^{-c L}$ for some model-dependent constant $c$. If such a segment exists, it contributes an excitation of energy $E \propto e^{-L/\xi}$, with $\xi$ the localization length of the MZMs. It has recently been appreciated though that the localization length of MZMs is strongly renormalized by the proximity coupling \cite{sankar_mzm_loclength, peng_mzm_loclength}. Since $\xi \propto \gamma^{-1}$, the required $L$ to produce a state of some fixed (low) energy decreases as $\gamma$ increases. The relative probability of generating this segment likewise increases, to the point where one (or more) of these states will reliably occur in the finite system for \emph{any} particular disorder potential. That is, as $\gamma$ is increased, the destruction of the topological gap and onset of a zero-energy DOS singularity, \emph{even for a short wire}, is driven by the exponentially increasing probability of making a domain that contributes a near-zero eigenvalue to the spectrum. In \fref{3}(b), we show the distribution of low-energy eigenvalues in a histogram over several disorder realizations, demonstrating this exponential dependence and its scaling with $\gamma$.

Of course, this has particularly unfortunate consequences for the application of wire edge MZMs in quantum information. Although a zero bias conductance peak may still arise from MZMs localized near the wire edges, the system is by no means topologically protected since there are many other MZMs localized at random spatial locations along the wire (in fact, the conductance signature should be similar to that found in \cite{flensberg_mbschain}). Indeed, these bulk low-energy modes will participate in braiding in an uncontrollable way. An upside, however, is that this situation cannot persist down to arbitrarily small $\gamma$ (again, in a finite wire), since we eventually return to the protected regime, where the TS gap is small ($\propto \gamma$) but unaffected by SC disorder.


\begin{figure}[t]
\centering
\includegraphics[width=0.48\textwidth]{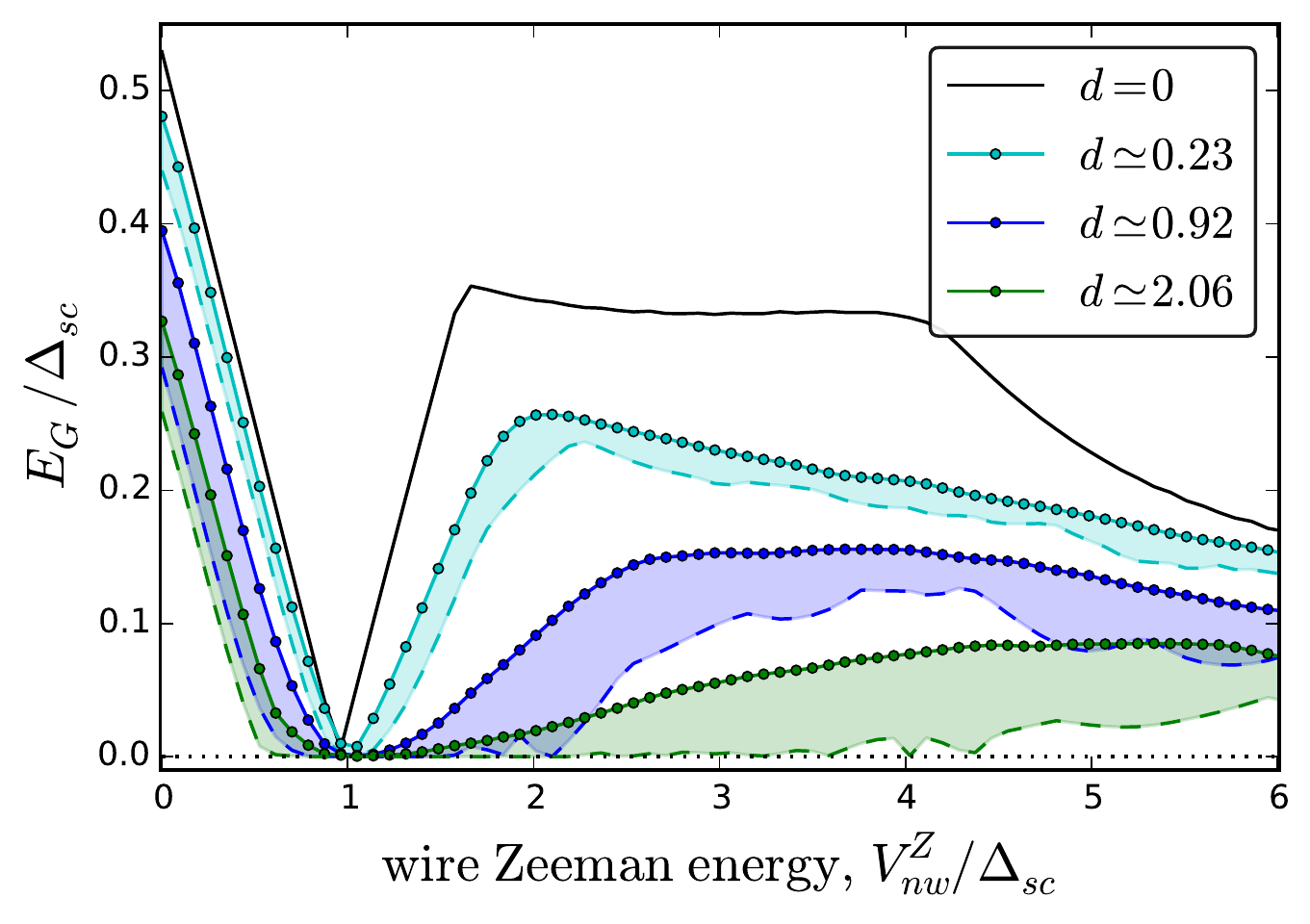}
\caption{
(Color online)
Spectral gap versus Zeeman energy in the nanowire for several values of disorder at fixed intermediate coupling $\gamma = \Delta_{sc}$. As in \fref{2}(a), the markers represent the disorder averaged $E_G$ while the dashed lines designate the minimum $E_G$ over all disorder samples. In addition to substantially suppressing the topological gap at $V^Z_{nw} > \gamma$, there is a large, disorder-induced gapless regime, such that a very large Zeeman energy is required to reopen the gap.
}
\flabel{4}
\end{figure}

{\bf Field-tuned topological phase transition:}
The previous section analyzed how disorder and strong proximity coupling conspire to suppress topological superconductivity. In experiments, however, $\gamma$ will be fixed, while the topological phase transition is tuned by an external magnetic field. In \fref{4} we show the disorder averaged $E_G$ at $\gamma = 1$ across this field-tuned phase transition. The dashed lines of \fref{4} again highlight the minimum $E_G$ over many disorder realizations. On the non-topological side $V^Z_{nw} < \gamma$ the gap suppression is rather mild, as in \fref{1}. While the average $E_G$ shows a ``rounded" approach to zero at the transition, the gap in the disorder averaged DOS, sensitive to the minimum $E_G$, can close well before the transition. This effect is far more substantial on the topological side where, for $d \simeq 2.06$, the gapless region persists up to $V^Z_{nw} \sim 4$. In all cases, however, at sufficiently large $V^Z_{nw}$ the DOS gap reopens and any wire edge MZMs are energetically isolated. However, large fields are undesirable as they suppress the parent SC, and also the TS gap is small in the high-field regime going approximately as $1/\sqrt{V^Z_{nw}}$.


{\bf Discussion:}
We conclude by summarizing the main message of this work. Strong proximity coupling is experimentally desirable for maximizing the induced pair correlations and spectral gap in the nanowire, however, this also necessarily eliminates the protection, predicted for weak coupling, of the induced superconductivity in the nanowire against SC disorder. Realistically, the parent SC materials in present experiments are quite disordered and the resulting induced disorder in the nanowire can easily be sufficient to close the spectral gap and, in particular, produce an accumulation of low-energy MZMs in the wire bulk, consistent with previously studied effective models of disordered semiconductor nanowires. New to this work is a technique addressing this issue directly, beyond effective models or Born approximation, which also characterizes the entire wire spectrum, rather than just the edge LDOS.
The expected benefits of strong proximity coupling vanish in the presence of SC disorder, even if the wire itself has zero disorder, suggesting that experiments should aim for the protected limit of $\gamma \ll \Delta_{sc}$. This also suggests the optimality of larger gap (but still BCS-like) parent superconductors, both to increase the overall energy scale and to decrease disorder effects through the shorter coherence length.
While our results are obtained for a quasi-one-dimensional model, the general principle extends just as well to synthetic TS in two dimensional SM-SC hybrids, where the experimental status is advancing rapidly \cite{expt_2dSMSC_arxiv1,expt_2dSMSC_arxiv2}.

This work is supported by Microsoft Q and LPS-MPO-CMTC.

\input{majorana_prox.bbl}


\clearpage


\begin{widetext}

\begin{center}
{\large Supplementary material for ``Proximity effect and Majorana bound states in clean semiconductor nanowires coupled to disordered superconductors"}
\label{supp}
\end{center}

\section{Model and methods}
The hamiltonian naturally decomposes into three parts: $H = H_{sc} + H_{nw} + T_{\gamma}$. The superconducting part of the system is a conventional $s$-wave superconductor, described by a square lattice BdG hopping model with on-site singlet pairing, Zeeman spin-splitting, and nonmagnetic potential disorder,
\be
\begin{aligned}
H_{sc} =
-t_{sc} &\sum_{\lr{ij}\sigma} \left( \adag_{i\sigma} \aaaa_{j\sigma} + \mbox{h.c.} \right)
+ \Delta_{sc} \sum_{i} \left( \adag_{i\su}\adag_{i\sd} - \aaaa_{i\su}\aaaa_{i\sd} \right) \\
+ &\sum_{i} \adag_{i} \left( \left[ 4t_{sc} - \mu_{sc} + W_{sc}(i) \right] \sigma_0 - E^Z_{sc} \sigma_x \right) \aaaa_{i} 
\end{aligned}
\ee
The nanowire has no intrinsic pairing, but includes a spin-orbit coupling $\propto k_x \sigma_y$,
\be
\begin{aligned}
H_{nw} =
-t_{nw} &\sum_{\lr{ij}\sigma} \left( \cdag_{i\sigma} \cccc_{j\sigma} + \mbox{h.c.} \right)
+ \alpha_{nw} \sum_{i} \left( \cdag_{i+\hat{x}} [i \sigma_y] \cccc_{i} + \mbox{h.c.} \right) \\
+ &\sum_{i} \cdag_{i} \left( \left[ 2t_{nw} - \mu_{nw} + W_{nw}(i) \right] \sigma_0 - E^Z_{nw} \sigma_x \right) \cccc_{i} 
\end{aligned}
\ee
\end{widetext}

To complete the model, we take a single chain of $H_{nw}$ sites and couple them to a ribbon of $H_{sc}$ sites through a nearest-neighbor interface term
\be
T_{\gamma} = -t_{\gamma} \sum_{i\sigma} \left( \cdag_{i\sigma} \aaaa_{i\sigma} + \mbox{h.c.} \right)
\ee
(The lattice indices are assumed the same for the NW and the first chain of SC sites; they are distinguished by their creation and annihilation operators.)

We are primarily interested in spectral features on the order $\Delta_{sc}$, so we take that as our natural unit of energy. In the next section we provide numerical values and justifications for all of the model parameters.

For all simulations we consider very large lattices with $N_s$ sites. Since each site has $4$ degrees of freedom (spin and particle-hole) there will be a total of $4 N_s$ states. This makes direct diagonalization unfeasible, but the hamiltonian is sparse -- there will be of order $N_s$, as opposed to $N_s^2$, nonzero matrix elements. Further, we expect relatively few of the eigenvalues to be on the scale of the parent gap, as this is, realistically, a \emph{small} energy scale (i.e., $\Delta_{sc} \ll t_{sc}, \mu_{sc}, t_{nw})$. In particular, we primarily wish to describe a region of the spectrum of width $\sim 2\Delta_{sc}$, while the difference between the largest and smallest eigenvalues of the full BdG matrix is set by the nanowire combined particle and hole bandwidth, at $\sim 8 t_{nw}$, and $t_{nw} \gg \Delta_{sc}$ (by two orders of magnitude). To get the eigenvalues and associated eigenvectors we apply the Lanczos method with the standard shift-and-invert scheme appropriate to interior eigenvalue problems, as implemented in ARPACK.

\begin{figure}[b]
\centering
\includegraphics[width=0.45\textwidth]{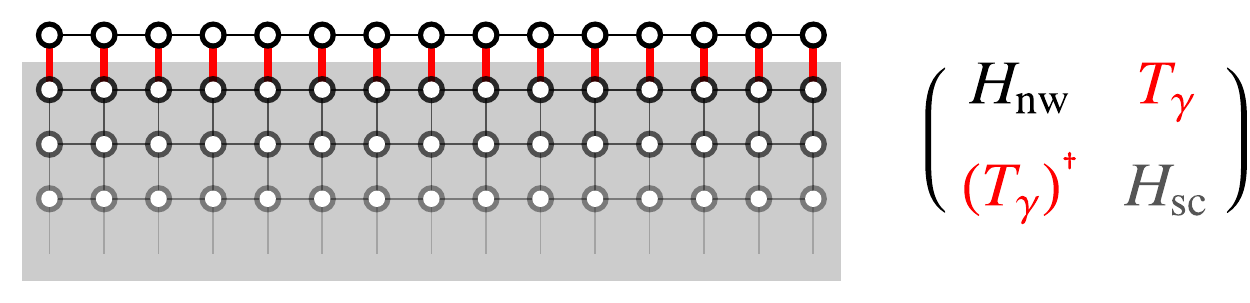}
\caption{Schematic representation of our lattice model. The grey region represents a superconducting bulk which is treated in the BdG formalism; $T_\gamma$ represents the nearest-neighbor tunneling matrix elements ($t_\gamma$) across the superconductor/semiconductor interface.}
\flabel{0}
\end{figure}

\section{Model parameters}

\subsection{Superconductor parameters}

We begin with the geometry. The superconductor is discretized on a two-dimensional square lattice, with lattice constant $a$. The dimensions are then $(L_x, L_y) = (N_x a, N_y a)$, and we use $N_x = 1500, N_y = 60$.

Next, $\Delta_{sc}$ is our unit of energy, and we describe a realistic separation of scales by taking $\mu_{sc} = 60 \Delta_{sc}$. This in turn determines an appropriate value for $t_{sc}$ -- we do not want the presence of the square lattice to significantly influence our results, so $\mu_{sc}$ should be sufficiently less than half the bandwidth $8t_{sc}$. To ensure that we are in that limit, we take $t_{sc} = \mu_{sc}/2 = 30 \Delta_{sc}$.

In principle this also sets our lattice parameter, since $a = \sqrt{\frac{\hbar^2}{2m^\ast_{sc} t_{sc}}}$ and the effective mass is realistically similar to the bare electron mass. Once $\Delta_{sc}$ is given in absolute units then this parameter is determined. Taking a value $\Delta_{sc} = 0.2$ meV appropriate to aluminum, for example, gives $a \sim 2.5$nm, or alternatively $N_x = 1500$ describes a $\sim 3.75 \mu$m wire, which is reasonable.

On the other hand, this simple model does not describe realistically the Fermi velocity or wavelength of \emph{aluminum} strictly speaking, however, the values are fine for representing the substantial separation in scales with corresponding values for the semiconductor nanowire, as discussed in the next subsection.

In both the wire and SC, the Zeeman energy is given in terms of the applied field $B$ as $E^Z_{\alpha} = \frac{1}{2} g_{\alpha} \mu_B B$. For a conventional superconductor, we take $g_{sc}  = 2$. The disorder potentials are $W_{\alpha}(i) \in [-w_\alpha, w_\alpha]$, i.e., at each site the potential is a uniformly chosen random value from the ``box" distribution of width $2w_\alpha$.

However, to make better contact with experiment and previous theoretical work, we need to reparametrize the SC disorder. The box distribution disorder that we implement microscopically is a convenient and widely used model, but the associated energy scale in the hamiltonian does not have any straightforward, experimentally meaningful analog. To make better connection to experiment and previous theoretical work, we estimate (from Fermi's golden rule) the associated mean free path $\lambda$ in the superconductor for a given parameter $w$ (the subscript $sc$ is implied in the following),
\be
\lambda = v_{F} \tau \simeq \frac{\hbar v_F}{2\pi} \left( a^2 \nu_{\rm 2D}(E_F) \frac{w^2}{3} \right)^{-1}
\ee
We then use the 2D density of states appropriate to the superconductor (in the normal state) and ignore small quantitative complications from the underlying square lattice,
\be
\nu_{\rm 2D}(E) = \frac{m^*}{\pi \hbar^2} \: \Rightarrow \: a^2 \nu_{\rm 2D}(E) = \frac{1}{2 \pi t},
\ee
Now, using the BCS coherence length $\xi = \frac{\hbar v_F}{\pi \Delta}$ (which corresponds to $\sim 27a$ for our choice of parameters), we can write down two dimensionless measures of disorder strength in terms of experimentally accessible length scales:
\be
\eta_1 \equiv \frac{\xi}{\lambda} = \frac{1}{3\pi} \frac{w^2}{\Delta t}
\ee
\be
\eta_2 \equiv \frac{1}{k_F \lambda} = \frac{1}{6} \frac{w^2}{\mu t}
\ee

Since $t = 30 \Delta$ and $\mu = 60 \Delta$, taking $w = t$ gives $\eta_1 \simeq 3.2$ and $\eta_2 = 1/12$. That is, the mean free path is much shorter than the SC coherence length, but still much longer than the SC Fermi wavelength. We therefore expect disorder in this regime to qualitatively reflect experiments.

Because we have fixed the other parameters, these two scales are not independent. However, following \cite{hoi_SCBA} we combine them into one dimensionless measure of disorder, $d = \sqrt{\eta_1 \eta_2}$. The values used in the paper are given in the following Table.

\setlength{\tabcolsep}{10pt}
\begin{center}
  \begin{tabular}{ c c c c }
    \hline
    $w/\Delta$ & $\eta_1$ & $\eta_2$ & $d$ \\ \hline
    20 & $\sim$1.41 & $\sim$0.037 & $\sim$0.23 \\
    40 & $\sim$5.66 & $\sim$0.148 & $\sim$0.92 \\
    60 & $\sim$12.7 & $\sim$0.333 & $\sim$2.06 \\
    \hline
  \end{tabular}
\end{center}

\subsection{Wire parameters}

We take a single chain to represent a single channel semiconductor nanowire, neglecting the higher-energy sub-bands. Electrons in the wire have a substantially smaller effective mass than in the SC. Assuming a shared lattice constant, we account for this qualitatively by setting $t_{nw} = 120 \Delta_{sc} > t_{sc}$, rather than using a realistic value of the effective mass. This is because $t_{nw}$ also sets the scale of finite size gaps $\sim t_{nw}/N_x^2$ which we want to keep small. We choose $\mu_{nw} = 0$ both for simplicity and to reflect the desired low carrier density in the wire.

The lattice spin orbit coupling scale is related to the physical Rashba velocity $\alpha_R$ as
\be
\alpha_{nw} = \frac{\hbar \alpha_{R}}{2a} = \sqrt{E_{s.o.} t_{nw}}, \quad E_{s.o.} = \frac{m^\ast_{nw} \alpha_R^2}{2}
\ee
and we choose a realistic value $\alpha_{nw} = 0.05 t_{nw} = 6 \Delta_{sc}$, corresponding to $E_{s.o.} = 0.3 \Delta_{sc}$

For the Zeeman energy, we choose a much larger g-factor in the wire, $g_{nw} = 20$. Finally, we only consider cases where the nanowire is completely clean, $W_{nw}(i)~=~0$.

\subsection{Proximity coupling parameter}
As with the disorder, to compare with prior work it will be useful to reparametrize the interface coupling, not as the microscopic tunneling matrix element $t_\gamma$ directly, but with an energy scale $\gamma \propto \left| t_{\gamma} \right|^2$. The proportionality factor should have units of energy$^{-1}$ and arises from the local density of superconductor states at the interface. We fit this disorder-dependent proportionality factor to recover the correct limiting behavior, $\Delta_{\rm ind} \simeq \gamma$ for $\gamma \ll \Delta$.

\section{The self-energy model}
For completeness, we include a formal description of the self-energy model - i.e., ``integrating out" the superconductor degrees of freedom. This is a completely generic approach and is manifestly exact for any quadratic model. Given a hamiltonian matrix $H$ describing the entire hybrid system, we can separate the Hilbert space into blocks describing spatial degrees of freedom in the wire ($\mathcal{A}$) and the superconductor ($\mathcal{B}$), as in \fref{0}. We otherwise leave the spatial indices implicit in the matrix structure of the following equations. The full resolvent matrix for this hamiltonian is
\be
G(\omega) = \begin{pmatrix} \omega - H_{\mathcal{A}} & \mathcal{T} \\ \mathcal{T}^\dagger & \omega - H_{\mathcal{B}} \end{pmatrix}^{-1}
\ee
and since we are interested only in the amplitudes for propagation between points in $\mathcal{A}$, we only need the upper left block of the inverted matrix. This is easily found to be
\be
G_{\mathcal{A}} (\omega) = \left( \omega - {H_\mathcal{A}} - \mathcal{T} \left( \omega - H_{\mathcal{B}} \right)^{-1} \mathcal{T}^{\dagger} \right)^{-1}
\ee
or, defining $G_{\alpha}^{(0)} \equiv \left( \omega - H_{\alpha} \right)^{-1}$, we can write
\be
[G_{\mathcal{A}} (\omega)]^{-1} =  [G^{(0)}_{\mathcal{A}} (\omega)]^{-1} - \mathcal{T} G^{(0)}_{\mathcal{B}} (\omega) \mathcal{T}^{\dagger}
\ee
which is formally a Dyson equation for the Green's function in region $\mathcal{A}$ with a self-energy matrix
\be
\Sigma(\omega) \equiv \mathcal{T} G^{(0)}_{\mathcal{B}} (\omega) \mathcal{T}^{\dagger}
\ee

To this point no approximations have been made. However, treating the induced self-energy exactly requires similar computational resources to diagonalizing the full hamiltonian $H$. One therefore typically relies on self-energies with simple analytic forms (i.e., a clean, infinite superconductor) or introduces some approximations (i.e., Born approximation to treat disorder in the superconductor) at this stage.

As an example, invoking the simple, well-justified, and frequently used self-energy model
\be
\Sigma(\omega) = -\gamma \left( \frac{\omega + \Delta \sigma_y \tau_y}{\sqrt{\Delta^2 - \omega^2}} \right) 
\ee
the Green's function for region $\mathcal{A}$ can be reorganized, and to linear order in $\omega/\Delta$ we have
\begin{align}
[G_{\mathcal{A}} (\omega \ll \Delta)]^{-1} &= \omega - H_\mathcal{A} - \Sigma(\omega) \\
& \simeq \left( 1 + \frac{\gamma}{\Delta} \right) \omega - H_\mathcal{A} + \gamma \sigma_y \tau_y \\
& = Z^{-1} \left( \omega - Z H_{\mathcal{A}} + \tilde{\Delta} \sigma_y \tau_y \right)
\end{align}
where $Z = \left( 1 + \frac{\gamma}{\Delta} \right)^{-1}$ renormalizes the ``bare" energy scales associated with $H_\mathcal{A}$, while $\tilde{\Delta} = Z\gamma = \frac{\gamma \Delta}{\gamma + \Delta}$ is the induced pair potential.

In this reduced model, for $H_{\mathcal{A}} = H_{nw}$, the condition for opening a topological gap becomes
\be
V^Z_{nw} > Z^{-1}\sqrt{(Z\mu_{nw})^2 + (Z\gamma)^2} = \sqrt{\mu_{nw}^2 + \gamma^2}
\ee
In other words, because of the downward renormalization of the bare wire energy scale, the bare Zeeman energy needs to exceed the coupling, rather than the induced gap, even in the strong coupling limit where $\gamma \gg \tilde{\Delta}$.

The superconducting coherence length in the nanowire should be obtained from the renormalized Fermi velocity and pair potential as
\be
\xi = \frac{\hbar (Z v_F^{nw})}{\pi(Z \gamma)} = \frac{\hbar v_F^{nw}}{\pi \gamma}
\ee
which suggests that the coherence length is independent of the spectral gap or the pair potential, and can be made arbitrarily short by increasing $\gamma$.

\section{Pair breaking from pairing inhomogeneity}
That a spatially varying pair potential gives rise to similar DOS behavior as magnetic disorder is easily demonstrated (and shown in \cite{takei_prl}), following the Abrikosov-Gorkov pair breaking formalism. Writing a generalized disorder potential
\be
\hat{U} = U_1(\vec{r}) \tau_z + U_2(\vec{r}) \vec{S} \cdot \vec{\alpha} + U_3(\vec{r}) \sigma_y \tau_y
\ee
where the last term has the same Nambu structure as the pairing, we note that the difference between magnetic and nonmagnetic impurities arises from the fact that
\be
\{\tau_z, \sigma_y \tau_y\} = 0, \quad
[\vec{S} \cdot \vec{\alpha}, \sigma_y \tau_y] = 0
\ee
for arbitrary $\vec{S}$. Obviously $[\sigma_y \tau_y, \sigma_y \tau_y] = 0$ as well, so the two potentials $U_2,U_3$ contribute in an identical fashion to the renormalized Green's function in Born approximation.

\end{document}

%% file: majorana_prox.bbl
%

%% file: majorana_prox.bbl
\begin{thebibliography}{30}%
\makeatletter
\providecommand \@ifxundefined [1]{%
 \@ifx{#1\undefined}
}%
\providecommand \@ifnum [1]{%
 \ifnum #1\expandafter \@firstoftwo
 \else \expandafter \@secondoftwo
 \fi
}%
\providecommand \@ifx [1]{%
 \ifx #1\expandafter \@firstoftwo
 \else \expandafter \@secondoftwo
 \fi
}%
\providecommand \natexlab [1]{#1}%
\providecommand \enquote  [1]{``#1''}%
\providecommand \bibnamefont  [1]{#1}%
\providecommand \bibfnamefont [1]{#1}%
\providecommand \citenamefont [1]{#1}%
\providecommand \href@noop [0]{\@secondoftwo}%
\providecommand \href [0]{\begingroup \@sanitize@url \@href}%
\providecommand \@href[1]{\@@startlink{#1}\@@href}%
\providecommand \@@href[1]{\endgroup#1\@@endlink}%
\providecommand \@sanitize@url [0]{\catcode `\\12\catcode `\$12\catcode
  `\&12\catcode `\#12\catcode `\^12\catcode `\_12\catcode `\%12\relax}%
\providecommand \@@startlink[1]{}%
\providecommand \@@endlink[0]{}%
\providecommand \url  [0]{\begingroup\@sanitize@url \@url }%
\providecommand \@url [1]{\endgroup\@href {#1}{\urlprefix }}%
\providecommand \urlprefix  [0]{URL }%
\providecommand \Eprint [0]{\href }%
\providecommand \doibase [0]{http://dx.doi.org/}%
\providecommand \selectlanguage [0]{\@gobble}%
\providecommand \bibinfo  [0]{\@secondoftwo}%
\providecommand \bibfield  [0]{\@secondoftwo}%
\providecommand \translation [1]{[#1]}%
\providecommand \BibitemOpen [0]{}%
\providecommand \bibitemStop [0]{}%
\providecommand \bibitemNoStop [0]{.\EOS\space}%
\providecommand \EOS [0]{\spacefactor3000\relax}%
\providecommand \BibitemShut  [1]{\csname bibitem#1\endcsname}%
\let\auto@bib@innerbib\@empty
\bibitem [{\citenamefont {Fu}\ and\ \citenamefont {Kane}(2008)}]{fu_kane_tsc}%
  \BibitemOpen
  \bibfield  {author} {\bibinfo {author} {\bibfnamefont {L.}~\bibnamefont
  {Fu}}\ and\ \bibinfo {author} {\bibfnamefont {C.~L.}\ \bibnamefont {Kane}},\
  }\href {\doibase 10.1103/PhysRevLett.100.096407} {\bibfield  {journal}
  {\bibinfo  {journal} {Phys. Rev. Lett.}\ }\textbf {\bibinfo {volume} {100}},\
  \bibinfo {pages} {096407} (\bibinfo {year} {2008})}\BibitemShut {NoStop}%
\bibitem [{\citenamefont {Sau}\ \emph {et~al.}(2010{\natexlab{a}})\citenamefont
  {Sau}, \citenamefont {Lutchyn}, \citenamefont {Tewari},\ and\ \citenamefont
  {Das~Sarma}}]{sau_rashba_2d_prl}%
  \BibitemOpen
  \bibfield  {author} {\bibinfo {author} {\bibfnamefont {J.~D.}\ \bibnamefont
  {Sau}}, \bibinfo {author} {\bibfnamefont {R.~M.}\ \bibnamefont {Lutchyn}},
  \bibinfo {author} {\bibfnamefont {S.}~\bibnamefont {Tewari}}, \ and\ \bibinfo
  {author} {\bibfnamefont {S.}~\bibnamefont {Das~Sarma}},\ }\href {\doibase
  10.1103/PhysRevLett.104.040502} {\bibfield  {journal} {\bibinfo  {journal}
  {Phys. Rev. Lett.}\ }\textbf {\bibinfo {volume} {104}},\ \bibinfo {pages}
  {040502} (\bibinfo {year} {2010}{\natexlab{a}})}\BibitemShut {NoStop}%
\bibitem [{\citenamefont {Zhang}\ \emph {et~al.}(2008)\citenamefont {Zhang},
  \citenamefont {Tewari}, \citenamefont {Lutchyn},\ and\ \citenamefont
  {Das~Sarma}}]{zhang_prl2008}%
  \BibitemOpen
  \bibfield  {author} {\bibinfo {author} {\bibfnamefont {C.}~\bibnamefont
  {Zhang}}, \bibinfo {author} {\bibfnamefont {S.}~\bibnamefont {Tewari}},
  \bibinfo {author} {\bibfnamefont {R.~M.}\ \bibnamefont {Lutchyn}}, \ and\
  \bibinfo {author} {\bibfnamefont {S.}~\bibnamefont {Das~Sarma}},\ }\href
  {\doibase 10.1103/PhysRevLett.101.160401} {\bibfield  {journal} {\bibinfo
  {journal} {Phys. Rev. Lett.}\ }\textbf {\bibinfo {volume} {101}},\ \bibinfo
  {pages} {160401} (\bibinfo {year} {2008})}\BibitemShut {NoStop}%
\bibitem [{\citenamefont {Sato}\ \emph {et~al.}(2009)\citenamefont {Sato},
  \citenamefont {Takahashi},\ and\ \citenamefont {Fujimoto}}]{sato_prl2009}%
  \BibitemOpen
  \bibfield  {author} {\bibinfo {author} {\bibfnamefont {M.}~\bibnamefont
  {Sato}}, \bibinfo {author} {\bibfnamefont {Y.}~\bibnamefont {Takahashi}}, \
  and\ \bibinfo {author} {\bibfnamefont {S.}~\bibnamefont {Fujimoto}},\ }\href
  {\doibase 10.1103/PhysRevLett.103.020401} {\bibfield  {journal} {\bibinfo
  {journal} {Phys. Rev. Lett.}\ }\textbf {\bibinfo {volume} {103}},\ \bibinfo
  {pages} {020401} (\bibinfo {year} {2009})}\BibitemShut {NoStop}%
\bibitem [{\citenamefont {Mourik}\ \emph {et~al.}(2012)\citenamefont {Mourik},
  \citenamefont {Zuo}, \citenamefont {Frolov}, \citenamefont {Plissard},
  \citenamefont {Bakkers},\ and\ \citenamefont {Kouwenhoven}}]{mourik-expt}%
  \BibitemOpen
  \bibfield  {author} {\bibinfo {author} {\bibfnamefont {V.}~\bibnamefont
  {Mourik}}, \bibinfo {author} {\bibfnamefont {K.}~\bibnamefont {Zuo}},
  \bibinfo {author} {\bibfnamefont {S.~M.}\ \bibnamefont {Frolov}}, \bibinfo
  {author} {\bibfnamefont {S.~R.}\ \bibnamefont {Plissard}}, \bibinfo {author}
  {\bibfnamefont {E.~P. A.~M.}\ \bibnamefont {Bakkers}}, \ and\ \bibinfo
  {author} {\bibfnamefont {L.~P.}\ \bibnamefont {Kouwenhoven}},\ }\href
  {\doibase 10.1126/science.1222360} {\bibfield  {journal} {\bibinfo  {journal}
  {Science}\ }\textbf {\bibinfo {volume} {336}},\ \bibinfo {pages} {1003}
  (\bibinfo {year} {2012})}\BibitemShut {NoStop}%
\bibitem [{\citenamefont {Das}\ \emph {et~al.}(2012)\citenamefont {Das},
  \citenamefont {Ronen}, \citenamefont {Most}, \citenamefont {Oreg},
  \citenamefont {Heiblum},\ and\ \citenamefont {Shtrikman}}]{das-expt}%
  \BibitemOpen
  \bibfield  {author} {\bibinfo {author} {\bibfnamefont {A.}~\bibnamefont
  {Das}}, \bibinfo {author} {\bibfnamefont {Y.}~\bibnamefont {Ronen}}, \bibinfo
  {author} {\bibfnamefont {Y.}~\bibnamefont {Most}}, \bibinfo {author}
  {\bibfnamefont {Y.}~\bibnamefont {Oreg}}, \bibinfo {author} {\bibfnamefont
  {M.}~\bibnamefont {Heiblum}}, \ and\ \bibinfo {author} {\bibfnamefont
  {H.}~\bibnamefont {Shtrikman}},\ }\href {http://dx.doi.org/10.1038/nphys2479}
  {\bibfield  {journal} {\bibinfo  {journal} {Nat Phys}\ }\textbf {\bibinfo
  {volume} {8}},\ \bibinfo {pages} {887} (\bibinfo {year} {2012})}\BibitemShut
  {NoStop}%
\bibitem [{\citenamefont {Deng}\ \emph {et~al.}(2012)\citenamefont {Deng},
  \citenamefont {Yu}, \citenamefont {Huang}, \citenamefont {Larsson},
  \citenamefont {Caroff},\ and\ \citenamefont {Xu}}]{deng-expt}%
  \BibitemOpen
  \bibfield  {author} {\bibinfo {author} {\bibfnamefont {M.~T.}\ \bibnamefont
  {Deng}}, \bibinfo {author} {\bibfnamefont {C.~L.}\ \bibnamefont {Yu}},
  \bibinfo {author} {\bibfnamefont {G.~Y.}\ \bibnamefont {Huang}}, \bibinfo
  {author} {\bibfnamefont {M.}~\bibnamefont {Larsson}}, \bibinfo {author}
  {\bibfnamefont {P.}~\bibnamefont {Caroff}}, \ and\ \bibinfo {author}
  {\bibfnamefont {H.~Q.}\ \bibnamefont {Xu}},\ }\href {\doibase
  10.1021/nl303758w} {\bibfield  {journal} {\bibinfo  {journal} {Nano Letters}\
  }\textbf {\bibinfo {volume} {12}},\ \bibinfo {pages} {6414} (\bibinfo {year}
  {2012})}\BibitemShut {NoStop}%
\bibitem [{\citenamefont {Finck}\ \emph {et~al.}(2013)\citenamefont {Finck},
  \citenamefont {Van~Harlingen}, \citenamefont {Mohseni}, \citenamefont
  {Jung},\ and\ \citenamefont {Li}}]{finck-expt}%
  \BibitemOpen
  \bibfield  {author} {\bibinfo {author} {\bibfnamefont {A.~D.~K.}\
  \bibnamefont {Finck}}, \bibinfo {author} {\bibfnamefont {D.~J.}\ \bibnamefont
  {Van~Harlingen}}, \bibinfo {author} {\bibfnamefont {P.~K.}\ \bibnamefont
  {Mohseni}}, \bibinfo {author} {\bibfnamefont {K.}~\bibnamefont {Jung}}, \
  and\ \bibinfo {author} {\bibfnamefont {X.}~\bibnamefont {Li}},\ }\href
  {\doibase 10.1103/PhysRevLett.110.126406} {\bibfield  {journal} {\bibinfo
  {journal} {Phys. Rev. Lett.}\ }\textbf {\bibinfo {volume} {110}},\ \bibinfo
  {pages} {126406} (\bibinfo {year} {2013})}\BibitemShut {NoStop}%
\bibitem [{\citenamefont {Churchill}\ \emph {et~al.}(2013)\citenamefont
  {Churchill}, \citenamefont {Fatemi}, \citenamefont {Grove-Rasmussen},
  \citenamefont {Deng}, \citenamefont {Caroff}, \citenamefont {Xu},\ and\
  \citenamefont {Marcus}}]{churchill-expt}%
  \BibitemOpen
  \bibfield  {author} {\bibinfo {author} {\bibfnamefont {H.~O.~H.}\
  \bibnamefont {Churchill}}, \bibinfo {author} {\bibfnamefont {V.}~\bibnamefont
  {Fatemi}}, \bibinfo {author} {\bibfnamefont {K.}~\bibnamefont
  {Grove-Rasmussen}}, \bibinfo {author} {\bibfnamefont {M.~T.}\ \bibnamefont
  {Deng}}, \bibinfo {author} {\bibfnamefont {P.}~\bibnamefont {Caroff}},
  \bibinfo {author} {\bibfnamefont {H.~Q.}\ \bibnamefont {Xu}}, \ and\ \bibinfo
  {author} {\bibfnamefont {C.~M.}\ \bibnamefont {Marcus}},\ }\href {\doibase
  10.1103/PhysRevB.87.241401} {\bibfield  {journal} {\bibinfo  {journal} {Phys.
  Rev. B}\ }\textbf {\bibinfo {volume} {87}},\ \bibinfo {pages} {241401}
  (\bibinfo {year} {2013})}\BibitemShut {NoStop}%
\bibitem [{\citenamefont {{Rokhinson}}\ \emph {et~al.}(2012)\citenamefont
  {{Rokhinson}}, \citenamefont {{Liu}},\ and\ \citenamefont
  {{Furdyna}}}]{rokhinson_frac_ac}%
  \BibitemOpen
  \bibfield  {author} {\bibinfo {author} {\bibfnamefont {L.~P.}\ \bibnamefont
  {{Rokhinson}}}, \bibinfo {author} {\bibfnamefont {X.}~\bibnamefont {{Liu}}},
  \ and\ \bibinfo {author} {\bibfnamefont {J.~K.}\ \bibnamefont {{Furdyna}}},\
  }\href {\doibase 10.1038/nphys2429} {\bibfield  {journal} {\bibinfo
  {journal} {Nat Phys}\ }\textbf {\bibinfo {volume} {8}},\ \bibinfo {pages}
  {795} (\bibinfo {year} {2012})}\BibitemShut {NoStop}%
\bibitem [{\citenamefont {Sau}\ \emph {et~al.}(2010{\natexlab{b}})\citenamefont
  {Sau}, \citenamefont {Lutchyn}, \citenamefont {Tewari},\ and\ \citenamefont
  {Das~Sarma}}]{sau_robust}%
  \BibitemOpen
  \bibfield  {author} {\bibinfo {author} {\bibfnamefont {J.~D.}\ \bibnamefont
  {Sau}}, \bibinfo {author} {\bibfnamefont {R.~M.}\ \bibnamefont {Lutchyn}},
  \bibinfo {author} {\bibfnamefont {S.}~\bibnamefont {Tewari}}, \ and\ \bibinfo
  {author} {\bibfnamefont {S.}~\bibnamefont {Das~Sarma}},\ }\href {\doibase
  10.1103/PhysRevB.82.094522} {\bibfield  {journal} {\bibinfo  {journal} {Phys.
  Rev. B}\ }\textbf {\bibinfo {volume} {82}},\ \bibinfo {pages} {094522}
  (\bibinfo {year} {2010}{\natexlab{b}})}\BibitemShut {NoStop}%
\bibitem [{\citenamefont {Stanescu}\ \emph {et~al.}(2010)\citenamefont
  {Stanescu}, \citenamefont {Sau}, \citenamefont {Lutchyn},\ and\ \citenamefont
  {Das~Sarma}}]{tudor_prox}%
  \BibitemOpen
  \bibfield  {author} {\bibinfo {author} {\bibfnamefont {T.~D.}\ \bibnamefont
  {Stanescu}}, \bibinfo {author} {\bibfnamefont {J.~D.}\ \bibnamefont {Sau}},
  \bibinfo {author} {\bibfnamefont {R.~M.}\ \bibnamefont {Lutchyn}}, \ and\
  \bibinfo {author} {\bibfnamefont {S.}~\bibnamefont {Das~Sarma}},\ }\href
  {\doibase 10.1103/PhysRevB.81.241310} {\bibfield  {journal} {\bibinfo
  {journal} {Phys. Rev. B}\ }\textbf {\bibinfo {volume} {81}},\ \bibinfo
  {pages} {241310} (\bibinfo {year} {2010})}\BibitemShut {NoStop}%
\bibitem [{\citenamefont {Takei}\ \emph {et~al.}(2013)\citenamefont {Takei},
  \citenamefont {Fregoso}, \citenamefont {Hui}, \citenamefont {Lobos},\ and\
  \citenamefont {Das~Sarma}}]{takei_prl}%
  \BibitemOpen
  \bibfield  {author} {\bibinfo {author} {\bibfnamefont {S.}~\bibnamefont
  {Takei}}, \bibinfo {author} {\bibfnamefont {B.~M.}\ \bibnamefont {Fregoso}},
  \bibinfo {author} {\bibfnamefont {H.-Y.}\ \bibnamefont {Hui}}, \bibinfo
  {author} {\bibfnamefont {A.~M.}\ \bibnamefont {Lobos}}, \ and\ \bibinfo
  {author} {\bibfnamefont {S.}~\bibnamefont {Das~Sarma}},\ }\href {\doibase
  10.1103/PhysRevLett.110.186803} {\bibfield  {journal} {\bibinfo  {journal}
  {Phys. Rev. Lett.}\ }\textbf {\bibinfo {volume} {110}},\ \bibinfo {pages}
  {186803} (\bibinfo {year} {2013})}\BibitemShut {NoStop}%
\bibitem [{\citenamefont {Chang}\ \emph {et~al.}(2015)\citenamefont {Chang},
  \citenamefont {Albrecht}, \citenamefont {Jespersen}, \citenamefont
  {Kuemmeth}, \citenamefont {Krogstrup}, \citenamefont {Nyg{\aa}rd},\ and\
  \citenamefont {Marcus}}]{cm_natnano}%
  \BibitemOpen
  \bibfield  {author} {\bibinfo {author} {\bibfnamefont {W.}~\bibnamefont
  {Chang}}, \bibinfo {author} {\bibfnamefont {S.~M.}\ \bibnamefont {Albrecht}},
  \bibinfo {author} {\bibfnamefont {T.~S.}\ \bibnamefont {Jespersen}}, \bibinfo
  {author} {\bibfnamefont {F.}~\bibnamefont {Kuemmeth}}, \bibinfo {author}
  {\bibfnamefont {P.}~\bibnamefont {Krogstrup}}, \bibinfo {author}
  {\bibfnamefont {J.}~\bibnamefont {Nyg{\aa}rd}}, \ and\ \bibinfo {author}
  {\bibfnamefont {C.~M.}\ \bibnamefont {Marcus}},\ }\href
  {http://dx.doi.org/10.1038/nnano.2014.306} {\bibfield  {journal} {\bibinfo
  {journal} {Nat Nano}\ }\textbf {\bibinfo {volume} {10}},\ \bibinfo {pages}
  {232} (\bibinfo {year} {2015})}\BibitemShut {NoStop}%
\bibitem [{\citenamefont {Brydon}\ \emph {et~al.}(2015)\citenamefont {Brydon},
  \citenamefont {Das~Sarma}, \citenamefont {Hui},\ and\ \citenamefont
  {Sau}}]{brydon_prb}%
  \BibitemOpen
  \bibfield  {author} {\bibinfo {author} {\bibfnamefont {P.~M.~R.}\
  \bibnamefont {Brydon}}, \bibinfo {author} {\bibfnamefont {S.}~\bibnamefont
  {Das~Sarma}}, \bibinfo {author} {\bibfnamefont {H.-Y.}\ \bibnamefont {Hui}},
  \ and\ \bibinfo {author} {\bibfnamefont {J.~D.}\ \bibnamefont {Sau}},\ }\href
  {\doibase 10.1103/PhysRevB.91.064505} {\bibfield  {journal} {\bibinfo
  {journal} {Phys. Rev. B}\ }\textbf {\bibinfo {volume} {91}},\ \bibinfo
  {pages} {064505} (\bibinfo {year} {2015})}\BibitemShut {NoStop}%
\bibitem [{\citenamefont {Nadj-Perge}\ \emph {et~al.}(2014)\citenamefont
  {Nadj-Perge}, \citenamefont {Drozdov}, \citenamefont {Li}, \citenamefont
  {Chen}, \citenamefont {Jeon}, \citenamefont {Seo}, \citenamefont {MacDonald},
  \citenamefont {Bernevig},\ and\ \citenamefont {Yazdani}}]{yazdani_science}%
  \BibitemOpen
  \bibfield  {author} {\bibinfo {author} {\bibfnamefont {S.}~\bibnamefont
  {Nadj-Perge}}, \bibinfo {author} {\bibfnamefont {I.~K.}\ \bibnamefont
  {Drozdov}}, \bibinfo {author} {\bibfnamefont {J.}~\bibnamefont {Li}},
  \bibinfo {author} {\bibfnamefont {H.}~\bibnamefont {Chen}}, \bibinfo {author}
  {\bibfnamefont {S.}~\bibnamefont {Jeon}}, \bibinfo {author} {\bibfnamefont
  {J.}~\bibnamefont {Seo}}, \bibinfo {author} {\bibfnamefont {A.~H.}\
  \bibnamefont {MacDonald}}, \bibinfo {author} {\bibfnamefont {B.~A.}\
  \bibnamefont {Bernevig}}, \ and\ \bibinfo {author} {\bibfnamefont
  {A.}~\bibnamefont {Yazdani}},\ }\href {\doibase 10.1126/science.1259327}
  {\bibfield  {journal} {\bibinfo  {journal} {Science}\ }\textbf {\bibinfo
  {volume} {346}},\ \bibinfo {pages} {602} (\bibinfo {year}
  {2014})}\BibitemShut {NoStop}%
\bibitem [{\citenamefont {Albrecht}\ \emph {et~al.}(2016)\citenamefont
  {Albrecht}, \citenamefont {Higginbotham}, \citenamefont {Madsen},
  \citenamefont {Kuemmeth}, \citenamefont {Jespersen}, \citenamefont
  {Nyg{\aa}rd}, \citenamefont {Krogstrup},\ and\ \citenamefont
  {Marcus}}]{cm_nature_exppro}%
  \BibitemOpen
  \bibfield  {author} {\bibinfo {author} {\bibfnamefont {S.~M.}\ \bibnamefont
  {Albrecht}}, \bibinfo {author} {\bibfnamefont {A.~P.}\ \bibnamefont
  {Higginbotham}}, \bibinfo {author} {\bibfnamefont {M.}~\bibnamefont
  {Madsen}}, \bibinfo {author} {\bibfnamefont {F.}~\bibnamefont {Kuemmeth}},
  \bibinfo {author} {\bibfnamefont {T.~S.}\ \bibnamefont {Jespersen}}, \bibinfo
  {author} {\bibfnamefont {J.}~\bibnamefont {Nyg{\aa}rd}}, \bibinfo {author}
  {\bibfnamefont {P.}~\bibnamefont {Krogstrup}}, \ and\ \bibinfo {author}
  {\bibfnamefont {C.~M.}\ \bibnamefont {Marcus}},\ }\href
  {http://dx.doi.org/10.1038/nature17162} {\bibfield  {journal} {\bibinfo
  {journal} {Nature}\ }\textbf {\bibinfo {volume} {531}},\ \bibinfo {pages}
  {206} (\bibinfo {year} {2016})}\BibitemShut {NoStop}%
\bibitem [{\citenamefont {Lutchyn}\ \emph {et~al.}(2012)\citenamefont
  {Lutchyn}, \citenamefont {Stanescu},\ and\ \citenamefont
  {Das~Sarma}}]{lutchyn_disorder}%
  \BibitemOpen
  \bibfield  {author} {\bibinfo {author} {\bibfnamefont {R.~M.}\ \bibnamefont
  {Lutchyn}}, \bibinfo {author} {\bibfnamefont {T.~D.}\ \bibnamefont
  {Stanescu}}, \ and\ \bibinfo {author} {\bibfnamefont {S.}~\bibnamefont
  {Das~Sarma}},\ }\href {\doibase 10.1103/PhysRevB.85.140513} {\bibfield
  {journal} {\bibinfo  {journal} {Phys. Rev. B}\ }\textbf {\bibinfo {volume}
  {85}},\ \bibinfo {pages} {140513} (\bibinfo {year} {2012})}\BibitemShut
  {NoStop}%
\bibitem [{\citenamefont {Potter}\ and\ \citenamefont
  {Lee}(2011)}]{potter_lee_erratum}%
  \BibitemOpen
  \bibfield  {author} {\bibinfo {author} {\bibfnamefont {A.~C.}\ \bibnamefont
  {Potter}}\ and\ \bibinfo {author} {\bibfnamefont {P.~A.}\ \bibnamefont
  {Lee}},\ }\href {\doibase 10.1103/PhysRevB.84.059906} {\bibfield  {journal}
  {\bibinfo  {journal} {Phys. Rev. B}\ }\textbf {\bibinfo {volume} {84}},\
  \bibinfo {pages} {059906} (\bibinfo {year} {2011})}\BibitemShut {NoStop}%
\bibitem [{\citenamefont {Lutchyn}\ \emph {et~al.}(2010)\citenamefont
  {Lutchyn}, \citenamefont {Sau},\ and\ \citenamefont
  {Das~Sarma}}]{lutchyn_wire}%
  \BibitemOpen
  \bibfield  {author} {\bibinfo {author} {\bibfnamefont {R.~M.}\ \bibnamefont
  {Lutchyn}}, \bibinfo {author} {\bibfnamefont {J.~D.}\ \bibnamefont {Sau}}, \
  and\ \bibinfo {author} {\bibfnamefont {S.}~\bibnamefont {Das~Sarma}},\ }\href
  {\doibase 10.1103/PhysRevLett.105.077001} {\bibfield  {journal} {\bibinfo
  {journal} {Phys. Rev. Lett.}\ }\textbf {\bibinfo {volume} {105}},\ \bibinfo
  {pages} {077001} (\bibinfo {year} {2010})}\BibitemShut {NoStop}%
\bibitem [{\citenamefont {Oreg}\ \emph {et~al.}(2010)\citenamefont {Oreg},
  \citenamefont {Refael},\ and\ \citenamefont {von Oppen}}]{oreg_wire}%
  \BibitemOpen
  \bibfield  {author} {\bibinfo {author} {\bibfnamefont {Y.}~\bibnamefont
  {Oreg}}, \bibinfo {author} {\bibfnamefont {G.}~\bibnamefont {Refael}}, \ and\
  \bibinfo {author} {\bibfnamefont {F.}~\bibnamefont {von Oppen}},\ }\href
  {\doibase 10.1103/PhysRevLett.105.177002} {\bibfield  {journal} {\bibinfo
  {journal} {Phys. Rev. Lett.}\ }\textbf {\bibinfo {volume} {105}},\ \bibinfo
  {pages} {177002} (\bibinfo {year} {2010})}\BibitemShut {NoStop}%
\bibitem [{\citenamefont {Anderson}(1959)}]{anderson_thm}%
  \BibitemOpen
  \bibfield  {author} {\bibinfo {author} {\bibfnamefont {P.~W.}\ \bibnamefont
  {Anderson}},\ }\href {\doibase
  http://dx.doi.org/10.1016/0022-3697(59)90036-8} {\bibfield  {journal}
  {\bibinfo  {journal} {J. Phys. Chem. Sol.}\ }\textbf {\bibinfo {volume}
  {11}},\ \bibinfo {pages} {26 } (\bibinfo {year} {1959})}\BibitemShut
  {NoStop}%
\bibitem [{\citenamefont {Motrunich}\ \emph {et~al.}(2001)\citenamefont
  {Motrunich}, \citenamefont {Damle},\ and\ \citenamefont
  {Huse}}]{motrunich_pwave}%
  \BibitemOpen
  \bibfield  {author} {\bibinfo {author} {\bibfnamefont {O.}~\bibnamefont
  {Motrunich}}, \bibinfo {author} {\bibfnamefont {K.}~\bibnamefont {Damle}}, \
  and\ \bibinfo {author} {\bibfnamefont {D.~A.}\ \bibnamefont {Huse}},\ }\href
  {\doibase 10.1103/PhysRevB.63.224204} {\bibfield  {journal} {\bibinfo
  {journal} {Phys. Rev. B}\ }\textbf {\bibinfo {volume} {63}},\ \bibinfo
  {pages} {224204} (\bibinfo {year} {2001})}\BibitemShut {NoStop}%
\bibitem [{\citenamefont {Hui}\ \emph {et~al.}(2015)\citenamefont {Hui},
  \citenamefont {Sau},\ and\ \citenamefont {Das~Sarma}}]{hoi_SCBA}%
  \BibitemOpen
  \bibfield  {author} {\bibinfo {author} {\bibfnamefont {H.-Y.}\ \bibnamefont
  {Hui}}, \bibinfo {author} {\bibfnamefont {J.~D.}\ \bibnamefont {Sau}}, \ and\
  \bibinfo {author} {\bibfnamefont {S.}~\bibnamefont {Das~Sarma}},\ }\href
  {\doibase 10.1103/PhysRevB.92.174512} {\bibfield  {journal} {\bibinfo
  {journal} {Phys. Rev. B}\ }\textbf {\bibinfo {volume} {92}},\ \bibinfo
  {pages} {174512} (\bibinfo {year} {2015})}\BibitemShut {NoStop}%
\bibitem [{\citenamefont {Chevallier}\ \emph {et~al.}(2013)\citenamefont
  {Chevallier}, \citenamefont {Simon},\ and\ \citenamefont
  {Bena}}]{chevallier_disorder_coupling}%
  \BibitemOpen
  \bibfield  {author} {\bibinfo {author} {\bibfnamefont {D.}~\bibnamefont
  {Chevallier}}, \bibinfo {author} {\bibfnamefont {P.}~\bibnamefont {Simon}}, \
  and\ \bibinfo {author} {\bibfnamefont {C.}~\bibnamefont {Bena}},\ }\href
  {\doibase 10.1103/PhysRevB.88.165401} {\bibfield  {journal} {\bibinfo
  {journal} {Phys. Rev. B}\ }\textbf {\bibinfo {volume} {88}},\ \bibinfo
  {pages} {165401} (\bibinfo {year} {2013})}\BibitemShut {NoStop}%
\bibitem [{\citenamefont {Sarma}\ \emph {et~al.}(2015)\citenamefont {Sarma},
  \citenamefont {Hui}, \citenamefont {Brydon},\ and\ \citenamefont
  {Sau}}]{sankar_mzm_loclength}%
  \BibitemOpen
  \bibfield  {author} {\bibinfo {author} {\bibfnamefont {S.~D.}\ \bibnamefont
  {Sarma}}, \bibinfo {author} {\bibfnamefont {H.-Y.}\ \bibnamefont {Hui}},
  \bibinfo {author} {\bibfnamefont {P.~M.~R.}\ \bibnamefont {Brydon}}, \ and\
  \bibinfo {author} {\bibfnamefont {J.~D.}\ \bibnamefont {Sau}},\ }\href
  {http://stacks.iop.org/1367-2630/17/i=7/a=075001} {\bibfield  {journal}
  {\bibinfo  {journal} {New Journal of Physics}\ }\textbf {\bibinfo {volume}
  {17}},\ \bibinfo {pages} {075001} (\bibinfo {year} {2015})}\BibitemShut
  {NoStop}%
\bibitem [{\citenamefont {Peng}\ \emph {et~al.}(2015)\citenamefont {Peng},
  \citenamefont {Pientka}, \citenamefont {Glazman},\ and\ \citenamefont {von
  Oppen}}]{peng_mzm_loclength}%
  \BibitemOpen
  \bibfield  {author} {\bibinfo {author} {\bibfnamefont {Y.}~\bibnamefont
  {Peng}}, \bibinfo {author} {\bibfnamefont {F.}~\bibnamefont {Pientka}},
  \bibinfo {author} {\bibfnamefont {L.~I.}\ \bibnamefont {Glazman}}, \ and\
  \bibinfo {author} {\bibfnamefont {F.}~\bibnamefont {von Oppen}},\ }\href
  {\doibase 10.1103/PhysRevLett.114.106801} {\bibfield  {journal} {\bibinfo
  {journal} {Phys. Rev. Lett.}\ }\textbf {\bibinfo {volume} {114}},\ \bibinfo
  {pages} {106801} (\bibinfo {year} {2015})}\BibitemShut {NoStop}%
\bibitem [{\citenamefont {Flensberg}(2010)}]{flensberg_mbschain}%
  \BibitemOpen
  \bibfield  {author} {\bibinfo {author} {\bibfnamefont {K.}~\bibnamefont
  {Flensberg}},\ }\href {\doibase 10.1103/PhysRevB.82.180516} {\bibfield
  {journal} {\bibinfo  {journal} {Phys. Rev. B}\ }\textbf {\bibinfo {volume}
  {82}},\ \bibinfo {pages} {180516} (\bibinfo {year} {2010})}\BibitemShut
  {NoStop}%
\bibitem [{\citenamefont {{Shabani}}\ \emph {et~al.}(2015)\citenamefont
  {{Shabani}}, \citenamefont {{Kjaergaard}}, \citenamefont {{Suominen}},
  \citenamefont {{Kim}}, \citenamefont {{Nichele}}, \citenamefont
  {{Pakrouski}}, \citenamefont {{Stankevic}}, \citenamefont {{Lutchyn}},
  \citenamefont {{Krogstrup}}, \citenamefont {{Feidenhans'l}}, \citenamefont
  {{Kraemer}}, \citenamefont {{Nayak}}, \citenamefont {{Troyer}}, \citenamefont
  {{Marcus}},\ and\ \citenamefont {{Palmstr{\o}m}}}]{expt_2dSMSC_arxiv1}%
  \BibitemOpen
  \bibfield  {author} {\bibinfo {author} {\bibfnamefont {J.}~\bibnamefont
  {{Shabani}}}, \bibinfo {author} {\bibfnamefont {M.}~\bibnamefont
  {{Kjaergaard}}}, \bibinfo {author} {\bibfnamefont {H.~J.}\ \bibnamefont
  {{Suominen}}}, \bibinfo {author} {\bibfnamefont {Y.}~\bibnamefont {{Kim}}},
  \bibinfo {author} {\bibfnamefont {F.}~\bibnamefont {{Nichele}}}, \bibinfo
  {author} {\bibfnamefont {K.}~\bibnamefont {{Pakrouski}}}, \bibinfo {author}
  {\bibfnamefont {T.}~\bibnamefont {{Stankevic}}}, \bibinfo {author}
  {\bibfnamefont {R.~M.}\ \bibnamefont {{Lutchyn}}}, \bibinfo {author}
  {\bibfnamefont {P.}~\bibnamefont {{Krogstrup}}}, \bibinfo {author}
  {\bibfnamefont {R.}~\bibnamefont {{Feidenhans'l}}}, \bibinfo {author}
  {\bibfnamefont {S.}~\bibnamefont {{Kraemer}}}, \bibinfo {author}
  {\bibfnamefont {C.}~\bibnamefont {{Nayak}}}, \bibinfo {author} {\bibfnamefont
  {M.}~\bibnamefont {{Troyer}}}, \bibinfo {author} {\bibfnamefont {C.~M.}\
  \bibnamefont {{Marcus}}}, \ and\ \bibinfo {author} {\bibfnamefont {C.~J.}\
  \bibnamefont {{Palmstr{\o}m}}},\ }\href@noop {} {\bibfield  {journal}
  {\bibinfo  {journal} {ArXiv e-prints}\ } (\bibinfo {year} {2015})},\ \Eprint
  {http://arxiv.org/abs/1511.01127} {arXiv:1511.01127 [cond-mat.mes-hall]}
  \BibitemShut {NoStop}%
\bibitem [{\citenamefont {{Kjaergaard}}\ \emph {et~al.}(2016)\citenamefont
  {{Kjaergaard}}, \citenamefont {{Nichele}}, \citenamefont {{Suominen}},
  \citenamefont {{Nowak}}, \citenamefont {{Wimmer}}, \citenamefont
  {{Akhmerov}}, \citenamefont {{Folk}}, \citenamefont {{Flensberg}},
  \citenamefont {{Shabani}}, \citenamefont {{Palmstr{\o}m}},\ and\
  \citenamefont {{Marcus}}}]{expt_2dSMSC_arxiv2}%
  \BibitemOpen
  \bibfield  {author} {\bibinfo {author} {\bibfnamefont {M.}~\bibnamefont
  {{Kjaergaard}}}, \bibinfo {author} {\bibfnamefont {F.}~\bibnamefont
  {{Nichele}}}, \bibinfo {author} {\bibfnamefont {H.~J.}\ \bibnamefont
  {{Suominen}}}, \bibinfo {author} {\bibfnamefont {M.~P.}\ \bibnamefont
  {{Nowak}}}, \bibinfo {author} {\bibfnamefont {M.}~\bibnamefont {{Wimmer}}},
  \bibinfo {author} {\bibfnamefont {A.~R.}\ \bibnamefont {{Akhmerov}}},
  \bibinfo {author} {\bibfnamefont {J.~A.}\ \bibnamefont {{Folk}}}, \bibinfo
  {author} {\bibfnamefont {K.}~\bibnamefont {{Flensberg}}}, \bibinfo {author}
  {\bibfnamefont {J.}~\bibnamefont {{Shabani}}}, \bibinfo {author}
  {\bibfnamefont {C.~J.}\ \bibnamefont {{Palmstr{\o}m}}}, \ and\ \bibinfo
  {author} {\bibfnamefont {C.~M.}\ \bibnamefont {{Marcus}}},\ }\href@noop {}
  {\bibfield  {journal} {\bibinfo  {journal} {ArXiv e-prints}\ } (\bibinfo
  {year} {2016})},\ \Eprint {http://arxiv.org/abs/1603.01852} {arXiv:1603.01852
  [cond-mat.mes-hall]} \BibitemShut {NoStop}%
\end{thebibliography}
